\documentclass[11pt]{article}
\usepackage{blois,epsfig}
\def\be{\begin{equation}}
\def\ee{\end{equation}}
\def\bea{\begin{eqnarray}}
\def\eea{\end{eqnarray}}

\begin{document}
\vspace*{4cm}
\title{LESSONS FROM WINDOWS ON THE UNIVERSE}

\author{P.J.E. PEEBLES }

\address{Joseph Henry Laboratories, Princeton University,\\
 Princeton NJ 08544, USA}

\maketitle\abstracts{I know better than to come between the experts here assembled and their research programs, so I confine these remarks to lessons to be drawn on the state of our subject from the histories of research in three Windows on the Universe:  cosmology, our extragalactic neighborhood, and life in other worlds.}

\section{Cosmology at Redshift Less than $10^{10}$ May Be Nearing Completion}

Cosmology and astronomy operate on the Tantalus Condition: you can look but not touch, observe but not experiment. That particularly limits study of the large-scale nature of the universe; we can only observe the tiny parts of spacetime into the past near our light cone and near our world line.  But technology allows a considerable variety of observations in these windows, and what is observed can be interpreted in a simple and close to convincing way, in the $\Lambda$CDM theory of what happened in the last factor of $10^{10}$ expansion of the universe. 

Briefly, standard $\Lambda$CDM assumes conventional local physics, general relativity theory, expansion from a hot big bang with initially small adiabatic gaussian near scale-invariant departures from homogeneity and isotropy, cold dark matter, Einstein's cosmological constant $\Lambda$, and negligible space curvature. Some of these elements were chosen because they were seen to offer a promising fit to improving measurements. They must be counted as free ``parameters,'' along with the six or seven  adjustable parameters  (depending on how you count) in $\Lambda$CDM. It is important therefore that this cosmology passes a considerably greater variety of tests that could have falsified it. (The tests are reviewed, for example, in the book {\it Finding the Big Bang}.\,\cite{FTBB}) $\Lambda$CDM is predictive. 

This need not mean $\Lambda$CDM contains all the  physics that will be needed for analyses of the next generation of cosmological tests. But we can be reasonably sure that if a better cosmology is called for it will describe a universe that looks much like  $\Lambda$CDM, because that is what is observed. It is easy to imagine adjustments. For example, cosmic strings used to be considered a  plausible trigger for the formation of galaxies as well as an arguably natural product of a more complete particle physics. Better observations rule out the former, but the latter still suggests cosmic strings may play a role in a more complete cosmology. There also is a good chance of learning something new in the dark sector.

\subsection{Dark Energy}

The standard cosmology includes Einstein's cosmological constant, $\Lambda$, or a component in the stress-energy tensor that acts like it. This is a natural-looking  description of the quantum vacuum energy except that, as widely lamented, the numerical value from the cosmological tests seems ridiculously small. Might we be able to live with $\Lambda = 0$, maybe a natural outcome of a better quantum physics of the vacuum? 

In general relativity the Lagrangian density for gravity is $R$ plus a constant (where $R$ is the Ricci scalar and the constant represents $\Lambda$). One can choose instead a function $f(R)$ without the constant that fits the present tests about as well as general relativity. It is conceivable that improved tests will be better fit by $f(R)$ than $R$ plus a constant. That would be a really deep advance, though arguably even more puzzling than general relativity with $\Lambda$. 

Also under discussion is the idea that the strongly nonlinear clustering of matter in galaxies and concentrations of galaxies may have an appreciable effect on the Friedmann-Lema\^\i tre equations that describe the general expansion of the universe,  maybe resembling the effect of $\Lambda$. This is based on analyses of Raychaudhuri's equations that relate the evolution of the shear, vorticity, density, pressure and expansion along the world line of a fluid element. Within the reasonable assumption of a continuous differentiable fluid, and apart from orbit crossings, these relations are identities in general relativity. But averaging them to get the global rate of expansion is exceedingly complicated, as one sees in the literature of elegant mathematical analyses of this deeply nonlinear physics. There is a simpler way.~\cite{SL,SF} In general relativity theory the mass concentrations in stars and galaxies produce small perturbations to spacetime. These small spacetime perturbations are readily analized in perturbation theory. That indicates the observed departures from homogeneity affect the expansion of the universe, but only in the fractional amount $\sim v^2\sim 10^{-6}$, where $v\sim 300$ km~s$^{-1}$ is a typical velocity on the scale of  galaxies. The effect is real but much too small to account for the evidence for $\Lambda$.

I conclude we had best learn to live with dark energy. How might we understand its value? The anthropic argument is that in an ensemble of universes we live in one that allows galaxies that last 10 Gyr or so, as would be accommodated by values of $\Lambda$ that include $\Lambda$CDM. This neatly accounts for the small but nonzero dark energy density, but it raises a new question: is this what ``really'' happened, or a story invented to save the phenomenon? (Perhaps the context for the meaning of ``really'' will become clearer as the theory improves.)

What is the nature of the dark energy? Detection of evolution of the dark energy density, as opposed to Einstein's constant $\Lambda$, could be an invaluable hint; the search for evidence of evolution merits community support. But I question the value of a big science mission to search for this effect until the community has become better prepared to deal with the result. Maybe ancillary benefits for the broader community would justify a big science dark energy mission, but in that case the mission needs a better name; we don't want to appear to be gulling the funding agencies.

\subsection{Dark Matter}

The most widely discussed alternative to dark matter is MOdified Newtonian Dynamics, MOND.\,\cite{MOND}  It offers an account of the scaling relation $v\propto L^{1/4}$ known as the Tully-Fisher relation between the spiral galaxy circular velocity $v$ and luminosity $L$, and the Faber-Jackson relation between the stellar velocity dispersion and luminosity of ellipticals. The relations are even tighter when one uses baryon mass instead of luminosity, and extend to baryon masses below what was established when MOND was introduced. That is, in its intended application MOND is predictive. This is reason to continue to bear MOND in mind, but I think not a serious argument against $\Lambda$CDM. Doing away with nonbaryonic dark matter faces what I count as a daunting challenge: show how a generalization of MOND can fit the cosmological tests that on the face of it quite systematically point to a larger mass in nonbaryonic dark matter than in baryons.

The technology of detection of trace effects of dark matter (other than its gravity) is opening windows on its nature. We know there is a subdominant hot dark matter component, the sea of neutrinos left from the hot Big Bang.  We have constraints from the cosmological tests on the properties of the other components --- it seems reasonable to suppose there is more than one --- but that leaves room for something more interesting than the exceedingly simple physics of $\Lambda$CDM. Where might we look for hints to better physics? I offer one example. Apart from the most luminous members of rich clusters of galaxies, the environments of galaxies of a given type, elliptical or spiral, are quite heterogeneous. Numerical simulations of $\Lambda$CDM indicate that large galaxies exchange considerable amounts of matter with their surroundings at redshifts less than unity. Thus it is curious that scaling relations, including those mentioned above, are little affected by environment. (Bernardi {\it et al.}~\cite{Bernardi} Fig. 3 is a striking illustration.) Complex processes generate scaling relations on smaller scales, and maybe that is happening here. Or maybe the scaling relations are clues to a better cosmology, maybe better physics in the dark sector. 

\begin{figure}[t]
\begin{center}
\includegraphics[angle=0,width=6.in]{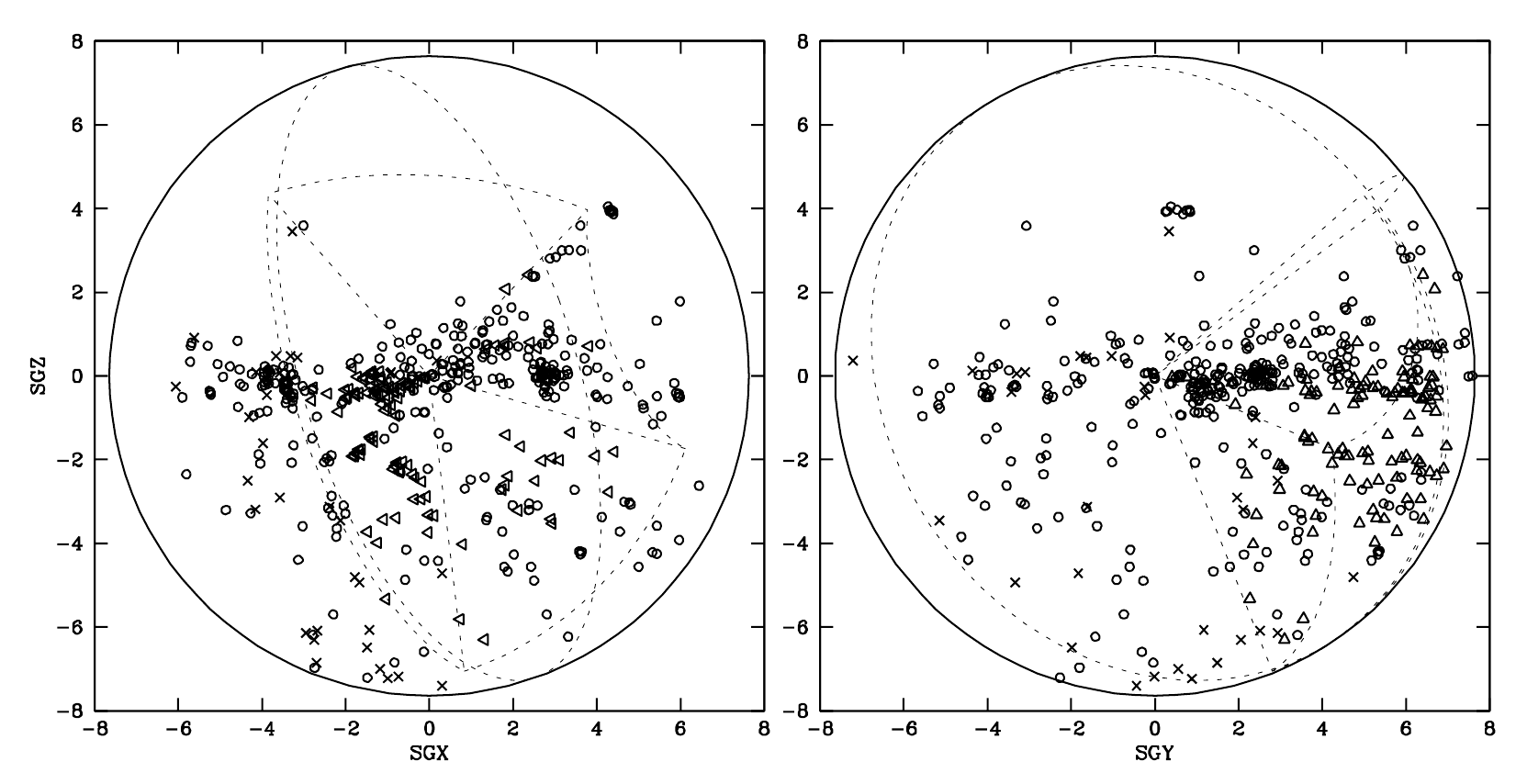} 
\caption{The galaxies at distances in the range $1<D<8$ Mpc identified on photographic plates (circles), in the Sloan Digital Sky Survey (triangles), and in the Parkes HI survey (crosses).\label{fig:1}}
\end{center}
\end{figure}

\section{Astronomy is Far From Complete}
\subsection{Our Extragalactic Neighborhood}

Another Window on the Universe is the nearest galaxies that can be observed in greatest detail. Figure~\ref{fig:1} shows the distribution of the known galaxies at distances $1<D<8$ Mpc. The lower bound excludes the members of the Local Group, which can be observed in even greater detail but may be  even further from a fair sample than what is shown here. The upper limit is 0.2\% of the Hubble Length, the distance at which Hubble's linear relation between recession velocity and distance extrapolates to the velocity of light.\footnote{More accurately, the upper limit in the figure is cosmological redshift 550 km~s$^{-1}$, which translates to a distance slightly less than 8~Mpc.} The circles show the 330 galaxies cataloged by Karachentsev {\it et al.}\,\cite{Kara}  Most were discovered on photographic plates, some from H{\small I} (21-cm atomic hydrogen line) searches for neighbors of known galaxies. The 115 added by the Sloan Digital Sky Survey are shown as triangles, the 35 added by the Parkes blind H{\small I} survey as crosses. The latter two surveys cover only parts of the sky, as roughly indicated by the  dotted lines. 

Harlow Shapley and Adelaide Ames, in 1932, remarked that many of the nearby galaxies appear in a band across the sky. Gerard de Vaucouleurs, in 1953, chose the supergalactic coordinates used in Figure~1 to put this planar concentration of the brighter galaxies along the supergalactic plane $SGZ=0$, with the $SGY$-axis pointing in the general direction of the Virgo Cluster of galaxies. It was later seen that the nearby fainter galaxies tend to be close to the supergalactic plane favored by more distant brighter ones. 

The faintest galaxies in this sample have luminosities $\sim 10^{-4}$ times that of the brightest. Still fainter dwarfs are known at $D<1$~Mpc; we can be sure deeper surveys will reveal many more at $D>1$~Mpc. Dwarfs further than about 0.3~Mpc from the nearest large galaxy tend to contain substantial atomic hydrogen in a rotationally supported gaseous disk or less commonly in irregular clouds around the stars. The 21-cm H{\small I} emission is readily  detectable though hard to find in the large space of sky and redshift. Dwarfs closer to a large galaxy tend to have much less H{\small I}, suggesting hydrogen disks are easily disturbed, changed to stars or plasma, by a large neighbor. But the big neighbor aids spotting these inconspicuous satellites.

The near empty Local Void at the upper left in the left-hand projection occupies about a third of the volume at $1<D<8$~Mpc. It contains just two of the 480 known galaxies.

The seven brightest galaxies are spirals, with a spread of luminosities of just a factor of two. Five are close to the supergalactic plane. M\,101 and NGC\,6946 are above plane, edging into the Local Void. NGC\,6946 is  the most isolated, with only a few small neighbors. Kormendy and Fisher\,\cite{Kormendy} note that NGC\,6946 is a near pure disc of stars, without the stellar bulge expected to result from mergers or close interactions with large neighbors after a disk of stars first formed. That seems consistent with its isolation. Andromeda has a bulge, likely therefore had a more active youth, perhaps in line with its more crowded environment. But its close neighbor, the Milky Way, is a near pure disk. Wyse\,\cite{Wyse} summarizes considerable other evidence that this galaxy has  suffered no significant disturbance since redshift $z\simeq 2$. The Milky Way and NGC\,6946 are not atypical: 11 of the 19 largest galaxies at $D<8$\,Mpc are near pure disks.\,\cite{Kormendy} 

This sample illustrates several of the properties of galaxies that do not seem to follow in a straightforward way from what is  seen in numerical simulations of structure formation in $\Lambda$CDM.

a. The galaxy luminosity function --- the PDF of galaxy luminosities --- has a sharp cutoff at the bright end. There are galaxies with stellar mass $\sim 10$ times that of the largest seven in this local sample, but they are exceedingly rare. In $\Lambda$CDM the mass function of dark matter halos that would be suitable homes for galaxies decreases much more slowly with increasing dark matter mass.  

b. The faint end of the galaxy luminosity function increases with decreasing luminosity distinctly less rapidly than the $\Lambda$CDM dark matter halo mass function increases with decreasing mass at the low mass end. 

c. Two of the seven largest galaxies, M\,101 and NGC\,6946, are in modest islands edging into the Local Void. But in simulations dark matter halos that are meet homes for spirals tend to avoid low density regions. 

d. NGC\,6946 does not have the bulge one associates with disturbances by massive neighbors after first disk formation, which seems consistent with its near isolation, but it is curious that it nevertheless was torqued up to a standard-looking angular momentum. 

e.  About half the 19 largest galaxies at $D<8$\,Mpc lack bulges, appearing to have evolved in close to isolation since first disk formation. This is not an argument against the hierarchical growth of structure --- mass concentrations gathering together to form larger concentrations --- but rather a question of timing of the major formation events, which seems out of synch with what is suggested by simulations.

f. The local void at $1<D<8$~Mpc contains just two of the known 480 galaxies, in about one third of the volume. In simulations voids are much less empty: the number density of dwarf halos in regions of low dark matter density is roughly 20\% of the global mean. Gas-rich dwarfs exist, and they avoid dense regions, so one might have thought they would be happy in voids. Why are they so rare in the Local Void? 

There are carefully considered proposals for how to understand these and other apparent anomalies in the theory and observations of galaxies. It is human nature to make up ``just so'' stories to account for anomalies; in the natural sciences they are termed ``working hypotheses.'' Are interpretations of the not inconsiderable list of apparent anomalies in the properties of galaxies well and truly established, or might they be closer to working hypotheses? Perhaps it's too soon to try to decide. 

I mention these issues not to carp or criticize but rather to make a point. Observations of the nearby galaxies that can be most closely studied bear on important issues in cosmology. The largest likely are all known, but as the observations improve we may expect continued debate on whether many of them really have evolved as island universes. At the time of writing we don't have  images of the distributions of atomic hydrogen in the two fascinating dwarfs in the Local Void, and we don't even have an image of the stars in one of them. There is the technology to do this, and to discover and examine many more galaxies at $1<D<8$~Mpc, maybe including some in the enigmatic Local Void. All these observations certainly will teach us something. I emphasize again, this is not a complaint, it's an illustration of the astronomers' long list of things to do. As they get to them what they find will more firmly establish or maybe drive adjustments of just so stories that seem to me to be suspiciously numerous.

\subsection{Life in Other Worlds}

A century ago Percival Lowell convinced himself that he could see canals on the planet Mars, maybe built by an advanced civilization that was running low on water. Astronomers at the time were skeptical of Lowell's observations, but the idea of canals on Mars certainly caught people's attention. I expect this is in part because it was a time of big civil engineering, including canals across Suez and later Panama. For the same reason I expect it is not entirely coincidental that in this digital age SETI is in effect looking for barcodes. The even newer age of analyses of the diversity of genetic codes was one path to the idea of a shadow biosphere (Davies {\it et al.}~\cite{Davies} and references therein) here on Earth, with a genetic code different from what can be isolated by the chemistry of established methods because it had a different origin with different chemistry. 

My point is that these are examples of socially-inspired research that is good science. (One may question Lowell's canals, but astronomers honor Lowell for building his observatory at a site that is good for astronomy and hiring good astronomers to instrument and use it to make capital discoveries, including the redshifts of galaxies.)  We may expect that the search for life in other worlds will continue to be inspired by what is happening in broader society, until discovery of a real phenomenon focusses attention. 

\section{Lessons}

What are the lessons to be drawn from these examples of research in Windows on the Universe? Here are my choices. 

1. {\em We are seeking reality.} Natural science operates on the hypothesis  that nature operates by rules that can be discovered, in successive approximations, by the interplay of theory and observation. This is a social construction. There are far better examples than cosmology of how spectacularly productive the program has been, but still I am deeply impressed that in the decade since $\Lambda$CDM became predictive it has passed increasingly demanding tests of what has happened on length scales more than ten orders of magnitude larger than the Solar System. 

2. {\em We won't complete the search.} I admire Weinberg's book,\,\cite{Dreams} {\em Dreams of a Final Theory} (by which is meant basic physics as opposed to its expression in complicated situations) but I suspect science  will become final in the sense that we are unable to do better. Cosmology at redshift $z>10^{10}$ may end as a theory that is consistent  internally and with all we know, adjustable to fit measurements as they come in, but not testable. 

3. {\em We are still making progress, but it's exceedingly uneven.} A half century ago the search for measures of the large-scale structure of the universe seemed exceedingly ambitious, but technology has led to a convincingly established cosmology. The search for life in other worlds has a much longer history, but the window hasn't opened yet. The physics of the dark sector of $\Lambda$CDM predicts that dark matter just piles up in halos while dark energy is constant or close to it. Physics in the visible sector is simple too, but capable of spectacularly complicated expression. Is the extreme simplicity of the dark sector only the easiest approximation we can get away with in a sector that has been only schematically explored? Returning to point 2, we may note that if $\Lambda$CDM physics differs from reality enough to matter it means it will be discovered, maybe in the laboratory, maybe in the astronomy. 

4. {\em Our progress is socially-shaped.} I risk reopening the science wars by offering a reminder of what I expect we've all experienced, that social forces help shape directions of research in the natural sciences. I mentioned examples in the search for life. I include examples in pure theory, as the anthropic argument. I include the influence of fashion, as in the relative attention paid to dark energy missions and our extragalactic neighborhood. This is an operating condition, and it can be productive, leading to motion toward point 1. 

5. {\em Our results are durable but not to be trusted.} Einstein wrote down general relativity theory nearly a century ago. His best test was the orbit of the planet Mercury. Now this theory passes demanding tests on the scale of the Hubble length, some $10^{15}$ times Mercury's distance from the Sun. This is spectacular durability. But a half century ago thoughtful physicists, including Einstein, presented good arguments against the cosmological constant. Now we learn we almost certainly have to live with it. The dark sector of $\Lambda$CDM passes searching tests. It is properly used as the basis for large-scale numerical simulations of structure formation, and for designs of observational programs and analyses of the results. But the dark sector physics is not to be trusted; it certainly could be found to be more interesting than $\Lambda$CDM.


\section*{References}
\begin{thebibliography}{99}

\bibitem{FTBB}P.J.E. Peebles, L.A. Page and R.B. Partridge,  {\em Finding the Big Bang} (Cambridge University Press, Cambridge, 2009).

 \bibitem{SL} U. Seljak and L. Hui, in {\em Clusters, Lensing,  and the Future of the Universe}, ed. V. Trimble and A. Reisenegger (Astronomical Society of the Pacific, San Francisco, 1996).

\bibitem{SF} E.~R. Siegel and J. N. Fry, {\em Astrophysical Journal Letters} {\bf 268}, L1 (2005).

\bibitem{MOND}  M. Milgrom, {\em Astrophysical Journal} {\bf 270}, 365 (1983).

\bibitem{Bernardi} M. Bernardi, R.C. Nichol, R.K. Sheth, C.J. Miller and J. Brinkmann, {\em Astronomical Journal} {\bf 131}, 1288 (2006).

\bibitem{Kara} I.D. Karachentsev, V.E. Karachentseva, W.K. Huchtmeier and D.I Makarov, {\em Astronomical Journal} {\bf 127}, 2031 (2004). 

\bibitem{Kormendy} J. Kormendy and D.B. Fisher, {\em  Astronomical Society of the Pacific Conference Series} {\bf 396}, 297 (2008).

\bibitem{Wyse} R.F.G. Wyse, {\em IAU Symposium} {\bf 258}, 11 (2009).

\bibitem{Davies} P.C.W. Davies, S.A. Benner, C.E. Cleleand, C.H. Lineweaver, C. P. McKay and F. Wolfe-Simon, {\em Astrobiology} {\bf 9}, 241 (2009). 

\bibitem{Dreams} S. Weinberg, {\em Dreams of a Final Theory} (Random House, New York, 1992).

\end{thebibliography}
\end{document}